\documentclass{elsart}
\usepackage{graphicx}
\usepackage{epsfig}
\newcommand{\bea}{\begin{equation}}
\newcommand{\eea}{\end{equation}}
\newcommand{\bean}{\begin{eqnarray}}
\newcommand{\eean}{\end{eqnarray}}
\def\L{\mathcal{L}}
\def\ni{\noindent}

\begin{document}
\begin{frontmatter}

\title{Lepton flavor violating Higgs boson decay $H\to \mu \tau$ at muon colliders}
\author[umsnh]{U. Cotti},
\author[umsnh]{M. Pineda}, and
\author[fcfmbuap]{G. Tavares-Velasco \thanksref{gtv}}
\address[umsnh]{Instituto de F\'\i sica y Matem\' aticas, Universidad
Michoacana de San Nicol\' as de Hidalgo, Apartado Postal 2-82, C. P.
58040, Morelia, Mich., M\' exico}
\address[fcfmbuap]{Facultad de Ciencias F\'\i sico Matem\'aticas, BUAP,
Apartado Postal 1152, 72000 Puebla, Pue. M\'exico }
\thanks [gtv] {E--mail: gtv@fcfm.buap.mx}


\begin{abstract}
We consider an effective nondiagonal coupling $H\,\bar{\mu}\,\tau$
and present the analysis of the Higgs boson mediated lepton-flavor
violating (LFV) reaction $\mu^-\mu^+\to\mu^{\pm}\tau^{\mp}$. For a
Higgs boson mass around 115 GeV and convenient values of the
strength of the coupling $H\,\bar{\mu}\,\tau$, which are within the
bounds obtained from the experimental limits on the LFV decays
$\tau^-\to\eta\mu^-$ and $\tau^-\to\mu^-\gamma$, we found that there
would be up to a few hundreds of $\mu^{\pm}\tau^{\mp}$ events per
year at a muon collider running with an integrated luminosity of 1
fb$^{-1}$. We paid special attention on the background for this LFV
reaction, which arises from the standard model process
$\mu^-\mu^+\to\mu^{\pm}\tau^{\mp}\bar{\nu}_{\mu}\nu_{\tau}$, and
discuss how it can be separated from the main signal.
\end{abstract}

\end{frontmatter}

Hopefully a Higgs boson will soon be detected, but it will still
remain to determine several of its properties, whose study may shed
light on the physics underlying the standard model (SM). One of the
main tasks of the present and future particle accelerators is thus
to perform a careful determination of the Higgs boson properties
such as mass, decay width, couplings to other  particles, and
properties under the discrete symmetries C and P. This will be the
goal of the CERN large hadron collider (LHC) but also of the next
generation colliders, for which there are several alternatives such
as a linear $e^-e^+$ collider and a muon collider. A major advantage
of a muon collider  is that it would operate as a Higgs boson
factory \cite{Ankenbrandt:1999as}, thereby offering great
opportunities for the study of the Higgs boson properties and
potential new physics effects in which this particle could play a
relevant role.

Since lepton flavor violation (LFV) is forbidden in the SM, any
observation of this class of effects would be a hint of new physics.
This has brought considerable attention to LFV. The prospect of a
muon collider, which would operate as a Higgs boson factory, opens
up the possibility for the study of scalar mediated LFV processes.
If the Higgs boson has nondiagonal couplings to the leptons, they
may become evident through the reaction $\mu^-\mu^+\to
H\to\mu^{\pm}\tau^{\mp}$. LFV mediated by a neutral  Higgs boson has
long been studied in specific models such as two-Higgs doublet
models (THDMs), suppersymmetry (SUSY) theories
\cite{Brignole:2003iv,Brignole:2004ah,Arganda:2004hz}, and other
beyond-the-SM scenarios \cite{Senami:2003jp}. Also, these
interactions were recently studied in a model independent way within
the framework of effective Lagrangians in Ref.
\cite{Diaz-Cruz:1999xe}. The possible detection of the decay
$H\to\mu \tau$ has already been considered at hadronic colliders
\cite{Han:2000jz}, muon colliders \cite{Sher:2000uq}, and $e^-e^+$
linear colliders \cite{Kanemura:2004cn}. As far as constraints on
the nondiagonal coupling $H\,\bar{l}_i\,l_j$ ($l_i=e,\,\mu,\,\tau$)
are concerned, they have been obtained from the LFV decays $l_i\to
l_j \bar{l}_k l_k$, $l_i\to l_j \gamma$, $l_i\to l_j \eta$, and the
muon anomalous magnetic moment \cite{Zhou:2001ew}. All these studies
have focused mainly on the most general THDM. In this work we will
present the study of LFV at a muon collider via
$\mu^-\mu^+\to\mu^\pm\tau^\mp$ scattering mediated by the Higgs
boson. The analysis will be performed within the framework of the
effective Lagrangian approach (ELA), which is tailored for the study
of new physics effects in a model independent fashion
\cite{Wudka:1994ny}.


The  $Hl_il_j$ coupling is induced by the following Yukawa-like
operator of dimension six \cite{Diaz-Cruz:1999xe}:

\bean O_{L_{\phi}}^{ij}= \phi^{\dagger}\phi\,\bar{L}_{i}^{'}\,\phi\,
l_{Rj}^{'}, \label{fax1} \eean where $L_{i}^{'}$ and $l_{Rj}^{'}$
represent the left-handed doublet and right-handed singlet of the
electroweak group, $\phi$  is the Higgs doublet and the subscripts
$i$ and $j$  stand for distinct lepton families, whereas the prime
denotes gauge eigenstates. Although the coupling $H\bar l_il_j$, and
also the $Z\bar l_il_j$ one, is induced at the tree-level by another
set of dimension six operators \cite{Flores-Tlalpa:2001sp}, the
contribution of such effective operators is suppressed by the factor
$m_{i,\,j}/m_Z$ and will be neglected from now on.

The effective operator (\ref{fax1}) induces the following Lorentz
structure for the LFV coupling $Hl_il_j$:

\bea \L^{Hl_{i}l_{j}}=\frac{i\,g\,\xi_{ij}}{2} \bar{l}_i\, l_j,
\label{fax12} \eea

\ni where $\xi_{ij}$ is an unknown coefficient that parametrizes our
ignorance of the new physics inducing LFV. Of course, when a
particular model is considered, $\xi_{ij}$ takes a particular form.
For instance, in the most general THDM \cite{Antaramian:1992ya},
dubbed model III, where scalar LFV couplings are allowed at the
tree-level, it is usual to consider the parametrization introduced
by Cheng and Sher \cite{Cheng:1987rs}: $\xi_{ij}=\lambda_{ij
}\sqrt{m_i m_j}/m_W$, where $\lambda_{ij}$ is a free parameter to be
constrained by low-energy experiments. This parametrization, which
is suited for models with multiple Higgs doublets, suggests that LFV
couplings involving the electron are naturally suppressed, whereas
LFV transitions involving the muon and the tau are much less
suppressed and can have a sizeable strength, which may be able to
give rise to effects that could be observed at particle colliders.
It has been suggested that $\lambda_{\mu\tau}\sim O(1)$
\cite{Cheng:1987rs}, though current constraints on this parameter
from experimental data are very weak and allow much larger values
for $\lambda_{\mu\tau}$ \cite{Zhou:2001ew}. Instead of considering a
specific model, we will pursue a model independent approach and
consider the most stringent bounds on $\xi_{\mu\tau}$ as obtained
from the most recent experimental data on LFV processes.

We will consider two possibilities for LFV in the Higgs sector. In
the SM there is only one Higgs doublet and the diagonalization of
the mass matrix simultaneously diagonalizes the matrix of Yukawa
couplings. LFV can arise when the Higgs sector is comprised by more
than one Higgs doublet or a more complex set of Higgs multiplets. In
the simplest case, the mere addition of only one Higgs doublet can
give rise to tree-level scalar LFV couplings such as occurs in the
model III. Nevertheless, it is well known  that any tree-level LFV
couplings of the Higgs boson can be eliminated by invoking a
discrete symmetry \cite{Glashow:1976nt}. In this scenario, this
class of effects can still arise at the one-loop level via the
virtual effects of new particles. A simple example of this scenario
is given by SUSY models, in which the LFV arises at the one-loop
level by the exchange of SUSY particles
\cite{Brignole:2003iv,Arganda:2004hz}. Instead of choosing a
particular model, effective Lagrangians allow one to study LFV in a
general fashion. In our analysis we will assume that there is a
relatively light Higgs boson whose behavior deviates  marginally
from that predicted by the SM, i.e. we will consider LFV effects
mediated by a SM-like Higgs boson.


In order to be able to make predictions it is necessary to give a
numerical value to the coefficient $\xi_{\mu\tau}$. According to the
effective Lagrangian philosophy, this parameter is to be bounded
from the current experimental limits on LFV processes such as the
decays $\tau^-\to \mu^-\mu^+\mu^-$, $\tau^-\to \mu^-\gamma$, and
$\tau^-\to \eta\mu^-$. The  branching ratio of the one-loop process
$\tau^-\to \mu^-\gamma$ reads, in the $m_\mu\to 0$ limit:

\bean Br(\tau^-\to \mu^-\gamma)=
\frac{\alpha^3\,|F(m_\tau,m_H)|^2\,m_\tau^3\,\xi_{\mu\tau}^2}
{512\,\pi^2\,s_{W}\,c_{W}\,m_Z^2\,\Gamma_{\tau}},
\eean

\ni where $\Gamma_{\tau}$ is the full $\tau$ width, and
$F(m_\tau,m_H)$ is given by

\bean F(m_\tau,m_H)&=&\frac{1}{2}+
\left(2-\frac{m_H^2}{m_\tau^2}\right)\left(B_0(m^2_\tau,m^2_\tau,m^2_H)-B_0(0,m^2_\tau,m^2_H)\right)
\nonumber\\&+&2\,m_\tau^2C_0(0,0,m_\tau^2,m_H^2,m_\tau^2,m_\tau^2).
\eean

\ni with $C_0$ and $B_0$ the usual Passarino-Veltman scalar
functions. The experimental bound on this decay is
\cite{Abe:2003sx}: $Br_{exp}(\tau^-\to \mu^-\gamma)\le 1.1\times
10^{-6}$.

As far as the decay $\tau^-\to \eta \mu^-$ is concerned, its
branching ratio is related to that of the decay
$\tau^-\to\mu^-\mu^+\mu^-$ as follows $Br(\tau^-\to \eta \mu^-)=8.4
Br(\tau^-\to\mu^-\mu^+\mu^-)$. Considering the leading term in
$m_\mu$ we have

\bea Br(\tau^-\to\mu^-\mu^+\mu^-)= \frac{1}{\Gamma_\tau}
\frac{\alpha^2\,m_\tau^5}{1536\,\pi\,s_W^4\,m_H^4}\left(\frac{m_\mu}{m_W}\right)^2
\xi_{\mu\tau}^2,\eea whereas the experimental limit on $\tau^-\to
\eta \mu^-$ is \cite{Enari:2004ax}: $Br_{exp}(\tau^-\to \eta
\mu^-)\le 3.4 \times 10^{-7}$.

The limits on $\xi_{\mu\tau}$ obtained via these decays are shown in
Table \ref{bounds} for different values of $m_H$. These values are
much less stringent than those that are obtained via the Cheng-Sher
ansatz, which is only appropriate for models with multiple Higgs
doublets \cite{Cheng:1987rs}. Nevertheless, our study is focused on
a broader class of models inducing LFV, including that class of
models in which the Cheng-Sher ansatz applies but also those
theories in which the LFV arise  at the one-loop level. Below we
will take a conservative approach and consider the values in the
range $10^{-3}$-$10^{-1}$ for $\xi_{\mu\tau}$.

\begin{table}[!hbt]
\begin{center}
\caption{\label{bounds} Bounds on the LFV parameter $\xi_{\mu\tau}$
as a function of the Higgs boson mass from the decays $\tau^-\to
\mu^-\mu^+\mu^-$ (first row) and $\tau^-\to \mu^-\gamma$ (second
row).}
\begin{tabular}{|c|c|c|c|c|c|}
\hline
$m_H$ (GeV)&120&130&140&150&200\\
\hline $\xi_{\mu\tau}\le$&$1.75$&$2.05$&
$2.38$&$2.73$&$4.87$\\
\hline $\xi_{\mu\tau}\le$&$1.61$&$1.85$&
$2.10$&$2.37$&$3.91$\\
\hline
\end{tabular}
\end{center}
\end{table}

We turn now to the calculation of scalar mediated
$\mu^-\mu^+\to\mu^{\pm}\tau^{\mp}$ scattering. We will neglect the
lepton masses everywhere except in the term associated with the
$H\bar\mu\mu$ coupling. It is straightforward to obtain the
unpolarized cross section after averaging over initial polarizations
and integrating over the scattering angle:

\bean\label{crossec} \sigma(\mu^-\mu^+\to
\mu^\pm\tau^\mp)=\frac{\pi^2\,\alpha^2\,m_\mu^2\,\xi_{\mu\tau}^2}{16\,s_W^4\,m_W^2\,\pi
s}(A_s+A_t+A_{st}), \eean with

\begin{figure}[!hbt]
\centering
\includegraphics[width=5in]{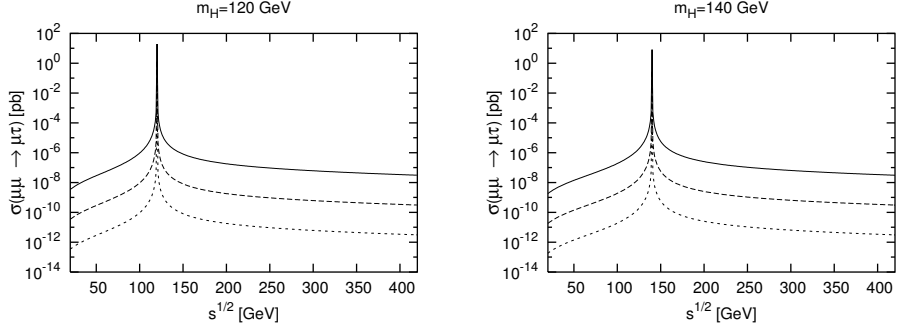}
\caption{\label{unpolarizedcs}Unpolarized
$\mu^-\mu^+\to\mu^\pm\tau^\mp$ cross section as a function of the
energy of the center of mass frame for two values of $m_H$ and
distinct values of $\xi_{\mu\tau}$: $10^{-1}$ (line), $10^{-2}$
(dashes) and $10^{-3}$ (points.)}
\end{figure}

\bean A_{s}&=&\frac{\hat s^2}{(\hat
s-1)^2+\hat\Gamma_H^2},\nonumber\\
A_{t}&=&\frac{2(1+\hat s)\log{\left(1+\hat s\right)}-\hat s(2+\hat
s)}{ \hat s(1+\hat
s)},\nonumber\\
A_{st}&=&\frac{\left(\hat s-1\right)\left(\hat{s}-\log{\left(1+\hat
s\right)}\right)\hat s}{(\hat s-1)^2+\hat\Gamma_H^2},\eean where
$\hat s=s/m_H^2$ and $\hat\Gamma_H=\Gamma_H/m_H$, with $s$ the
square of the center of mass energy of the muon collider, and
$\Gamma_H$ the  total Higgs boson decay width. In Fig.
\ref{unpolarizedcs} we show the numerical evaluation of
(\ref{crossec}) for different values of $\xi_{\mu\tau}$.  We have
assumed that $\Gamma_H$ is approximately given by the total decay
width of the SM Higgs boson, which was evaluated via the {\small
HDECAY} program \cite{Djouadi:1997yw}. As expected, we can see that
the $\mu^-\mu^+\to\mu^{\pm}\tau^{\mp}$ cross section is only
relevant in the resonance region, where it takes the familiar form:

\bea \sigma(\mu^-\mu^+\to \mu^\pm\tau^\mp)=\frac{4\pi}{m_H^2}
Br(H\to \mu^-\mu^+)Br(H\to \mu^\pm\tau^\mp), \eea with \bea
Br(H\rightarrow{\mu^-\mu^+})=\frac{\alpha\,m_{H}\,m_{\mu}^2}{8\,
s_W^2\,m_{W}^2\,\Gamma_H},\eea and \bea
Br(H\rightarrow{\mu^\pm\tau^\mp})=
\frac{\alpha\,m_{H}\,\xi_{\mu\tau}^2}{8s_W^2 \Gamma_H}.\eea

\begin{figure}[hbt!]
\centering
\includegraphics[width=4in]{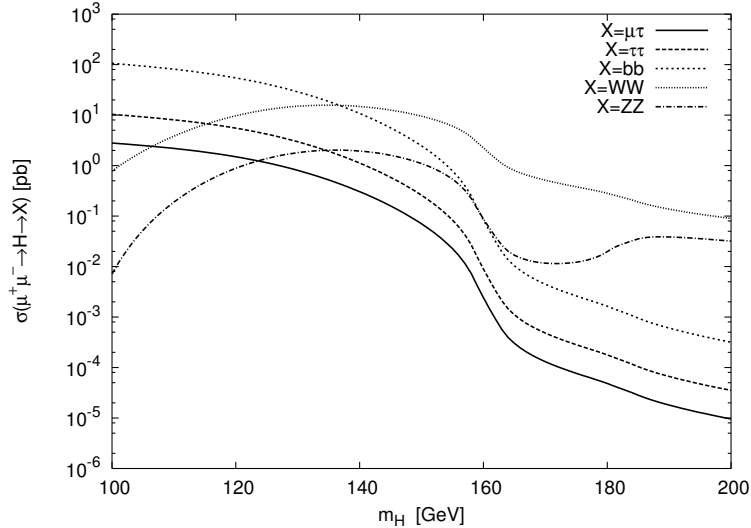}
\caption{\label{resonance} Unpolarized cross section for
$\mu^-\mu^+\to H\to X$ as a function of $m_H$ at a muon collider
running as a Higgs boson factory. For $m_H< 2\,m_V$ ($V=W,\,Z$), the
curves for $X=VV$ correspond to the production of an on-shell $V$
boson accompanied by a virtual one.}
\end{figure}

We will now concentrate on the scenario in which the muon collider
operates as a Higgs boson factory. In Fig. \ref{resonance} we show
the cross section for $\mu^-\mu^+\to\mu^\pm\tau^\mp$ scattering as a
function of the Higgs boson mass. Although we only show the
$\mu^-\mu^+\to\mu^\pm\tau^\mp$ cross section for
$\xi_{\mu\tau}=10^{-2}$, which is a moderate value if we consider
the bounds given in Table \ref{bounds}, the corresponding curve is
only shifted upwards (downwards) for larger (smaller) values of this
parameter. For comparison purposes, we also include the most
important decay channels of the Higgs boson. A future muon collider
is expected to work with an integrated luminosity of about 1
fb$^{-1}$ \cite{Ankenbrandt:1999as}. From Figure \ref{resonance} we
conclude that there would be up to a few hundred of
$\mu^\pm\tau^\mp$ events in a year for a Higgs boson with a mass
ranging between 100 and 140 GeV. For a heavier Higgs boson, the
cross section drops dramatically as more decay channels ($H\to WW$
and $H\to ZZ$, with both particles on-shell) become opened. These
results are in agreement with those presented in Ref.
\cite{Sher:2000uq} for the case of the THDM-III.

\begin{figure}[hbt!]
\centering
\includegraphics[width=3.5in]{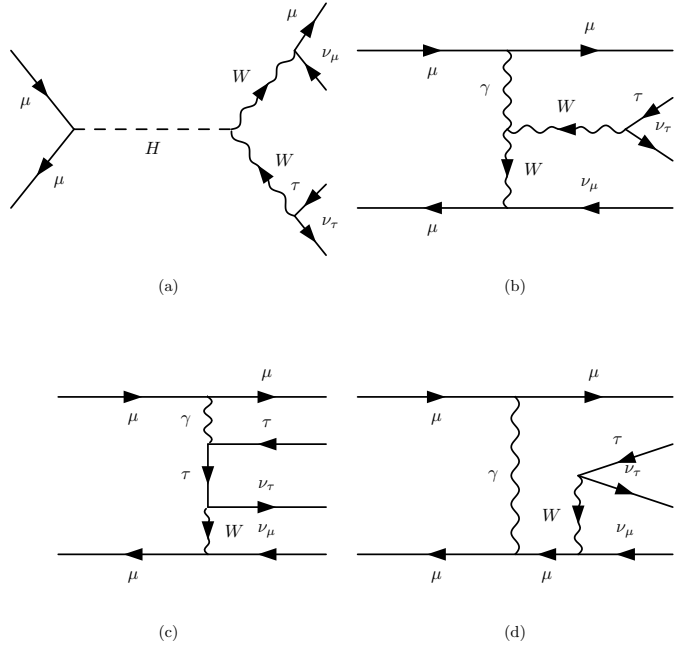}
\caption{\label{feyndiagbg} The dominant contributions to the SM
background  $\mu^-\tau^+\to \mu^\pm\tau^\mp
\bar\nu_{\mu}\nu_{\tau}$.}
\end{figure}

\begin{figure}[hbt!]
\centering
\includegraphics[width=3.5in]{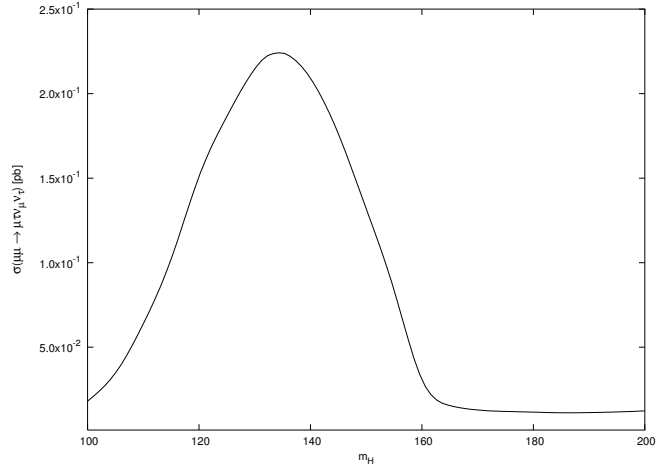}
\caption{\label{backgroundcs} Cross section for $\mu^-\tau^+\to
\mu^\pm\tau^\mp \bar\nu_{\mu}\nu_{\tau}$ scattering at
$\sqrt{s}=m_H$.}
\end{figure}

It is worthwhile to examine carefully the potential background for
the LFV process $\mu^-\mu^+\to \mu^\pm\tau^\mp$. In this reaction
the final leptons always emerge back to back and carrying a constant
energy which is one half the center of mass energy. The main
background would arise from the SM process $\mu^-\mu^+\to
\mu^\pm\tau^\mp \bar{\nu}_{\mu}\nu_\tau$, whose signature is a pair
$\mu^\pm\tau^\mp$ plus missings. There are 24 Feynman diagrams
contributing to this process, with exchange of the photon, the $Z$
boson, the $W$ boson and the Higgs boson itself. We have explicitly
calculated the contribution of these diagrams via the {\small
CALCHEP} package \cite{Pukhov:2004ca}. To discard those final
leptons emerging outside the detector coverage, we imposed the
following cut $|\cos\theta|\le 0.99$, where $\theta$ is the
scattering angle. It turns out that on the Higgs boson resonance the
dominant contribution comes from the diagrams shown in Fig.
\ref{feyndiagbg}, whereas the remaining diagrams give a negligible
contribution. For instance, those diagrams in which the photon and
the $Z$ boson are exchanged in the $s$ channel are suppressed by an
inverse factor of $m_H^4$ and $(m_H^2-m_Z^2)^2$, respectively. On
the other hand, the cross section of the diagrams in which the
photon couples directly to the initial and final muons is inversely
proportional to $m_H^4\,\sin^4(\theta/2)$ when the muon mass is
neglected and so it is considerably suppressed by the cut
$|\cos\theta|\le 0.99$. In these diagrams the final muon emerges
predominantly with low energy. Therefore, the main contribution to
the background comes from the diagram with Higgs boson exchange
[Fig. \ref{feyndiagbg} (a)]. The cross section arising from the
background process is shown in Fig. \ref{backgroundcs} as a function
of $m_H$. It reaches a peak around $m_H=130$ GeV, where it may be
larger than the LFV cross section, and then drops quickly.

\begin{figure}[hbt!]
\centering
\includegraphics[width=4in]{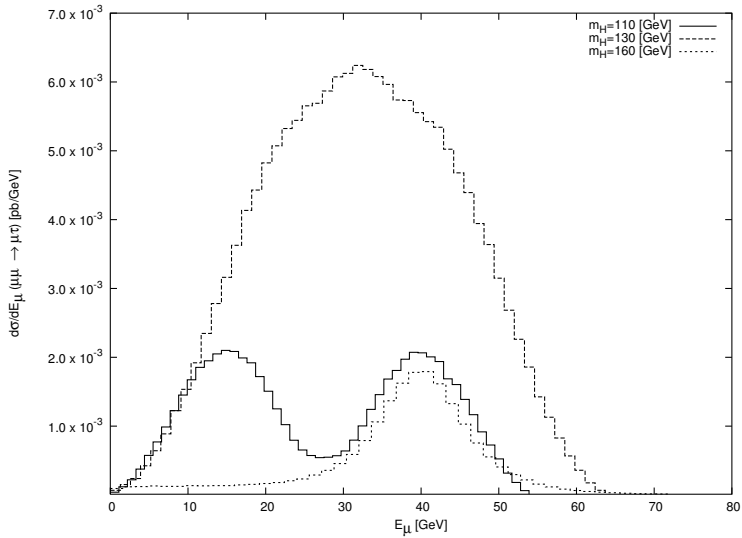}
\caption{\label{distribution} Energy distribution of the final muon
in $\mu^-\tau^+\to  \mu^\pm\tau^\mp \bar\nu_{\mu}\nu_{\tau}$
scattering for $\sqrt{s}=m_H$ and three values of $m_H$. The muons
emerging from the LFV process $\mu^-\tau^+\to \mu^\pm\tau^\mp$ are
monoenergetic, with an energy half the Higgs boson mass, so their
energy distribution exhibits a sharp peak  at $m_H/2$.}
\end{figure}

\begin{figure}[hbt!]
\centering
\includegraphics[width=4in]{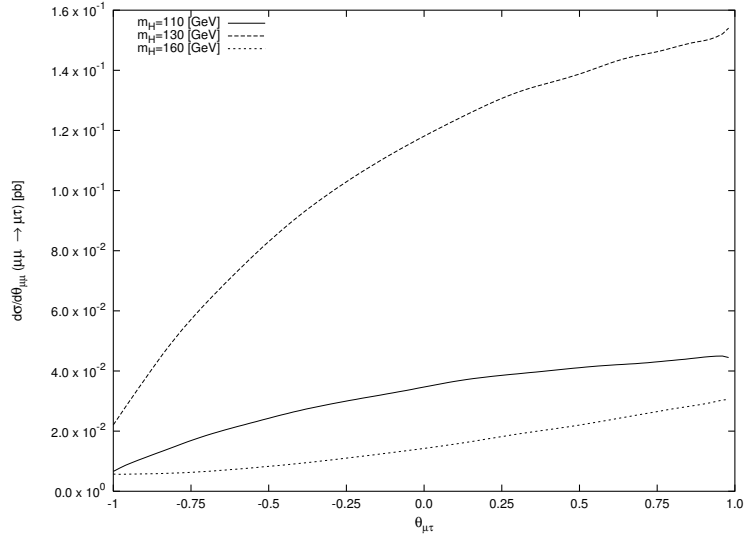}
\caption{\label{angdistribution} $\cos\theta_{\mu\tau}$ distribution
of $\mu^-\tau^+\to \mu^\pm\tau^\mp \bar\nu_{\mu}\nu_{\tau}$
scattering for $\sqrt{s}=m_H$ and three values of $m_H$. The $\mu$
and $\tau$ from the LFV process emerge back to back and so its
$\cos\theta_{\mu\tau}$ distribution exhibits a sharp peak at
$\cos\theta_{\mu\tau}=-1$.}
\end{figure}

We turn to some kinematical distributions that may be helpful to
separate the background from the signal. Since the background signal
is a $2\to 4$ process, the energy distribution of the final $\mu$
and $\tau$ would be essentially different from what would observed
in the $2\to 2$ reaction $\mu^-\tau^+\to \mu^\pm\tau^\mp$: while the
energy distribution of the $\mu$ and $\tau$ leptons emerging from
the latter process is peaked at $\sqrt{s}/2=m_H/2$, the $\mu$ and
$\tau$ emerging from the background process have a smaller and
nonuniform energy, which reach its maximal value at $m_H/2$. The
respective distribution is illustrated in Fig. \ref{distribution}
for various values of $m_H$, where we considered the dominant
contributions to $\mu^-\tau^+\to \mu^\pm\tau^\mp
\bar\nu_{\mu}\nu_{\tau}$. We can thus impose a hard cut on the
energy of the final particles and get rid of most of the background
events. We also show in Fig. \ref{angdistribution} the
$\theta_{\mu\tau}$ distribution, with $\theta_{\mu\tau}$ the angle
between the spatial momenta of the $\mu$ and $\tau$ emerging from
the background, for three values of $m_H$. This distribution is
peaked at $\cos\theta_{\mu\tau}=-1$ in the LFV  process. We can see
that the $\cos\theta_{\mu\tau}$ distribution arising from the
background is very different to that of the LFV signal. These and
other kinematical distributions along with appropriate cuts can be
used to separate the signal from the background. Furthermore, the
use of polarized beams can be helpful to separate those
contributions coming essentially from Higgs boson exchange to those
coming from other sources such as an extra $Z'$ boson.

We have presented  an analysis of lepton flavor violation within the
effective Lagrangian approach. We have considered the reaction
$\mu^-\tau^+\to \mu^\pm\tau^\mp$, which could be at the reach of a
future muon collider. We only have considered the scenario in which
the LFV arises from the Higgs boson.  For the strength of the
effective coupling $H\bar{\mu}\tau$, we assumed some values that are
within the constraints obtained from the LFV decays $\tau^-\to \mu^-
\gamma$ and $\tau^-\to\mu^-\eta$. From our analysis we can conclude
that a future muon collider may be useful to detect LFV mediated by
the Higgs boson. Our study also shows that the main background could
be separated from the signal through appropriate cuts and the study
of some kinematical distributions.

\ack{We acknowledge support from SNI and SEP-PROMEP (M\'exico).
G.T.V. thanks J.J. Toscano  for fruitful discussions. }


\begin{thebibliography}{99}

\bibitem{Ankenbrandt:1999as}
C.~M.~Ankenbrandt {\it et al.},
Phys.\ Rev.\ ST Accel.\ Beams {\bf 2} (1999) 081001; A.~Chao {\it et
al.},
in {\it Proc. of the APS/DPF/DPB Summer Study on the Future of
Particle Physics (Snowmass 2001) } ed. N.~Graf, eConf {\bf C010630}
(2001) MT1001; D.~B.~Cline and G.~G.~Hanson,
in {\it Proc. of the APS/DPF/DPB Summer Study on the Future of
Particle Physics (Snowmass 2001) } ed. N.~Graf, eConf {\bf C010630}
(2001) M103.


\bibitem{Brignole:2003iv}
A.~Brignole and A.~Rossi,
Phys.\ Lett.\ B {\bf 566} (2003) 217.

\bibitem{Brignole:2004ah}
A.~Brignole and A.~Rossi,
Nucl.\ Phys.\ B {\bf 701} (2004) 3.

\bibitem{Arganda:2004hz}
E.~Arganda, A.~M.~Curiel, M.~J.~Herrero and D.~Temes,
arXiv:hep-ph/0411048;
arXiv:hep-ph/0407302.

\bibitem{Senami:2003jp}
M.~Senami and K.~Yamamoto,
Phys.\ Rev.\ D {\bf 69} (2004);
W.~Grimus and L.~Lavoura,
Phys.\ Rev.\ D {\bf 66} (2002) 014016.

\bibitem{Diaz-Cruz:1999xe}
J.~L.~Diaz-Cruz and J.~J.~Toscano,
Phys.\ Rev.\ D {\bf 62} (2000) 116005.


\bibitem{Han:2000jz}
T.~Han and D.~Marfatia,
Phys.\ Rev.\ Lett.\  {\bf 86} (2001) 1442; U.~Cotti, L.~Diaz-Cruz,
C.~Pagliarone and E.~Vataga,
in {\it Proc. of the APS/DPF/DPB Summer Study on the Future of
Particle Physics (Snowmass 2001) } ed. N.~Graf, eConf {\bf C010630}
(2001) P102; K.~A.~Assamagan, A.~Deandrea and P.~A.~Delsart,
Phys.\ Rev.\ D {\bf 67} (2003) 035001.

\bibitem{Sher:2000uq}
M.~Sher,
Phys.\ Lett.\ B {\bf 487} (2000) 151.

\bibitem{Kanemura:2004cn}
S.~Kanemura, K.~Matsuda, T.~Ota, T.~Shindou, E.~Takasugi and
K.~Tsumura,
Phys.\ Lett.\ B {\bf 599} (2004) 83.


\bibitem{Zhou:2001ew}
Y.~F.~Zhou and Y.~L.~Wu,
Eur.\ Phys.\ J.\ C {\bf 27} (2003) 577;
Phys.\ Rev.\ D {\bf 64} (2001) 115018; S.~K.~Kang and K.~Y.~Lee,
Phys.\ Lett.\ B {\bf 521} (2001) 61; R.~A.~Diaz, R.~Martinez and
J.~A.~Rodriguez,
Phys.\ Rev.\ D {\bf 67} (2003) 075011; S. Nie and M. Sher and Phys.
Rev. D {\bf{58}}, 097701 (1998); M.~Kakizaki, Y.~Ogura and F.~Shima,
Phys.\ Lett.\ B {\bf 566} (2003) 210.

\bibitem{Wudka:1994ny}
J.~Wudka,
Int.\ J.\ Mod.\ Phys.\ A {\bf 9}(1994) 2301.

\bibitem{Flores-Tlalpa:2001sp}
A.~Flores-Tlalpa, J.~M.~Hernandez, G.~Tavares-Velasco and
J.~J.~Toscano,
Phys.\ Rev.\ D {\bf 65} (2002) 073010.

\bibitem{Antaramian:1992ya}
A.~Antaramian, L.~J.~Hall and A.~Rasin,
Phys.\ Rev.\ Lett.\  {\bf 69} (1992) 1871;
L.~J.~Hall and S.~Weinberg,
Phys.\ Rev.\ D {\bf 48} (1993) 979;
M.~E.~Luke and M.~J.~Savage,
Phys.\ Lett.\ B {\bf 307} (1993) 387.


\bibitem{Cheng:1987rs}
T.~P.~Cheng and M.~Sher,
Phys.\ Rev.\ D {\bf 35} (1987) 3484.

\bibitem{Glashow:1976nt}
S.~L.~Glashow and S.~Weinberg,
Phys.\ Rev.\ D {\bf 15} (1977) 1958.

\bibitem{Abe:2003sx}
K.~Abe {\it et al.},
Phys.\ Rev.\ Lett.\  {\bf 92} (2004) 171802.

\bibitem{Enari:2004ax}
Y.~Enari {\it et al.},
Phys.\ Rev.\ Lett.\  {\bf 93} (2004) 081803.

\bibitem{Djouadi:1997yw}
A.~Djouadi, J.~Kalinowski and M.~Spira,
Comput.\ Phys.\ Commun.\  {\bf 108} (1998) 56.

\bibitem{Pukhov:2004ca}
A.~Pukhov,
arXiv:hep-ph/0412191.


\end{thebibliography}
\end{document}